\documentstyle[12pt,psfig]{article}
\def\be{\begin{equation}}
\def\ee{\end{equation}}
\def\bea{\begin{eqnarray}}
\def\eea{\end{eqnarray}}

\def\a{\alpha}
\def\b{\beta}

\def\d{\partial}
\def\e{\epsilon}           % Also, \varepsilon
\def\g{\gamma}

\def\m{\mu}
\def\n{\nu}

\def\t{\tau}

\def\L{\Lambda}

\def\widebar{\overline}
\def\th{\theta}

\def\calL{{\cal L}}

\def\chidag{\chi^\dag}

\def\psib{\widebar{\psi}}
\def\eb{\widebar{\epsilon}}

\def\T{{\rm T}}
\def\M{{\rm M}}

\def\E{{\rm E}}

\def\dslash{\partial\!\!\! /  }
\def\Vslash{V\!\!\!\! /  }

\def\Dslash{D\!\!\!\! / \ }

\def\psidag{\psi^\dagger}
\def\chidag{\chi^\dagger}
\def\psibar{\widebar{\psi}}
\def\epstr{\epsilon^\top}
\def\psitr{\psi^\top}
\def\neta{\eta}

\def\inv{^{-1}}
\def\CE{C_\E}
\def\L{{\rm L}}
\def\R{{\rm R}}

\begin{document}

\thispagestyle{empty}
\begin{flushright}
{\sc ITP-SB}-96-44
\end{flushright}
\vspace{1cm}
\setcounter{footnote}{0}
\begin{center}
{\LARGE{On Euclidean spinors and Wick rotations
        \footnote{This research was supported in part by NSF grant no
 PHY9309888.}
    }}\\[14mm]

\sc{Peter van Nieuwenhuizen\footnote{e-mail:vannieu@insti.physics.sunysb.edu}
and Andrew Waldron\footnote{e-mail:wally@insti.physics.sunysb.edu},}\\[5mm]
{\it Institute for Theoretical Physics\\
State University of New York at Stony Brook\\
Stony Brook, NY 11794-3840, USA}\\[20mm]

{\sc Abstract}\\[2mm]
\end{center}

We propose a continuous Wick rotation for Dirac, Majorana and Weyl spinors
from Minkowski spacetime to Euclidean space which treats fermions on the
same footing as bosons. 
The result is a recipe to construct a supersymmetric Euclidean
theory from any supersymmetric Minkowski theory.
This Wick rotation is identified as a complex Lorentz boost
in a five-dimensional space and acts uniformly on bosons and fermions.
For Majorana and Weyl spinors our approach is 
reminiscent of the traditional Osterwalder Schrader approach in which
spinors are ``doubled'' but the action is not hermitean. 
However, for Dirac spinors our work provides a link to the work of
Schwinger and Zumino in which hermiticity is maintained but
spinors are not doubled. Our work differs from recent work by
Mehta since we introduce no external metric and 
transform only the basic fields.

\vfill

\newpage

\section{Introduction.}

Euclidean quantum field theory for bosons is used in particle physics
to justify certain manipulations in Minkowski quantum field theory, such
as the claim that for $t\rightarrow\pm\infty$ only the vacuum contributes
to the transition amplitude~\cite{Archimedes}.
 (Sometimes a term $i\epsilon\phi^2$
is added to the action by hand to obtain the same result 
in Minkowski spacetime, but this approach is somewhat ad hoc~\cite{Abers}).
Furthermore Euclidean field theory is crucial for the physics of instantons.
If one couples fermions to instantons, one needs Euclidean field theory
for spinors. In fact, an $N=2$ supersymmetric field theory has been
constructed in Euclidean space~\cite{Zumino}.
Also finite temperature field theory and lattice gauge theory\cite{Seiler}
use Euclidean field theory.
In Fujikawa's approach~\cite{Fujikawa} to anomalies in quantum field theories
one must regulate the Jacobian with a Gaussian integral whose part quadratic
in
momenta should be converging in all directions. To find the subleading terms
(for example, terms with $A_\mu\g_5$ or $iA_\m\g_5$) one would like to
know the continuation of the field theory to Euclidean space.
Some mathematical applications of recent work by Seiberg and Witten
are also based on supersymmetric theories continued to 
Euclidean space~\cite{Witten}.
In this article we present a new interpretation of the Wick 
rotation to Euclidean field theory. To avoid misunderstanding, we stress
that the object of our study is \underline{not} the Wick rotation
of the momentum variable $k_0$ in a Feynman amplitude; rather, we are
interested in the Wick rotation of the underlying field theory.

For spinors a quantum field theory in Euclidean space whose Green's
functions coincide with those of Minkowski spacetime after analytic 
continuation in the time coordinate, was constructed by Osterwalder
and Schrader (OS) in 1973~\cite{OS}.
Their work was based on canonical quantization rather
than path integrals and is, as a consequence, rather complicated. For 
example (i) a spinor field in four dimensional Euclidean space is expanded
in creation and annihilation operators depending on 
Euclidean \underline{four}-momenta $k_\mu^\E$, rather than three momenta
$\vec{k}$ as in the Minkowski case, (ii) fields do not (and cannot) 
obey field equations,
(iii) instead of the familiar spinors
$u_\alpha^{r}(\vec{k})$ and $v_\alpha^{r}(-\vec{k})$ (where $r=\pm$
and $\a=1,..,4$ is the spinor index) of 
Minkowski spacetime, one now has spinors $U^R_\alpha(k_\mu^\E)$,
$V^R_\alpha(-k_\mu^\E)$, $X^R_\alpha(k_\mu^\E)$ and $W^R_\alpha(-k_\mu^\E)$
where $R=1,..,4$.

However, if one rephrases their work in the language of path integrals,
then the only new aspect is that one should ``double'' the number of
spinor fields in Euclidean space. For example, for a Dirac spinor
the action in Minkowski spacetime and in Euclidean space, respectively, reads
\bea
\calL_\M=-\psidag i\gamma^0(\gamma^\mu\partial_\mu+m)\psi &;&
\{\gamma^\mu,\gamma^\nu\}=2\neta^{\mu\nu}\ \ ; \ \ \mu,\nu=0,..,3 \\
\calL_\E=-\chidag_\E (\gamma^\mu_\E\partial_\mu+m)\psi_\E \ \ \ &;&
\{\gamma^\mu_\E,\gamma^\nu_\E\}=2\delta^{\mu\nu}\ \ ; \ \ \mu,\nu=1,..,4\ 
\label{euclid}
\eea
where
$\chi^\dagger_\E$ 
is not related to $\psi_{{\rm E}}$ by complex or hermitean 
conjugation. We have written the word ``double''
in quotation marks because the number of Grassmann integration variables
is the same: $\psi$ and $\psidag$ in one case, and $\psi$ and $\chidag$
in the other\footnote{Notice that
in the path integral $\chi^\dagger$ and $\psi$ are treated as independent
variables while in the canonical approach they contain the same
(doubled) creation and annihilation operators. This is just like $\psi$ and
$\psidag$ in Minkowski space.}. The only property that has changed is 
hermiticity\footnote{
For anti-commuting variables we define 
$(\chi^\dagger\psi)^\dagger=\psidag\chi$. Our convention for the
metric signature in Minkowski spacetime is $\neta_{\mu\nu}=$
 ${\rm diag}(-,+,+,+)_{\mu\nu}$.}. The 
Euclidean action of OS is not hermitean. 

There are other approaches to Euclidean spinors, 
due to Schwinger~\cite{Schwinger} and Zumino~\cite{Zumino};
Fubini, Hanson and Jackiw~\cite{Jackiw}; and 
more recently Mehta~\cite{Mehta}.
Nicolai~\cite{Nicolai} extends the work of OS
to Majorana spinors\footnote{For a similar approach in two dimensions 
see~\cite{Borthwick}.} and defines the following action in Euclidean
space
\be
\calL_\E({\rm Maj})=-\frac{1}{2}\psitr_\E C (\gamma^\mu_\E\partial_\mu+
m)\psi_\E ,\label{Ni}
\ee
where $\psi_\E$ no longer satisfies a reality condition
and $C$ is the charge conjugation matrix which can equally well
be defined in Minkowski and Euclidean space (it satisfies
$C^\top =-C$ and $C\g_\mu C\inv =-\g_\mu^\top$). 
In Minkowski spacetime
one imposes the condition $\psidag i\gamma^0=\psitr C$ which yields a
hermitean action, but again $\calL_\E({\rm Maj})$ is not hermitean.
Zumino required that the Euclidean action be hermitean (which is not
necessary on physical grounds) and since no Euclidean
Majorana spinors exist (by which we mean spinors in Euclidean space
satisfying a reality condition), he considered an $N=2$ supersymmetric model
with Dirac spinors, whose kinetic term for the spinor fields coincides
with the original hermitean action proposed by Schwinger. Fubini et al.
use ``radial quantization'', in which the radius in Euclidean space plays 
the r\^{o}le of the time coordinate. (In string theory one usually continues
the world-sheet time $t$ to $-i\t$, and makes then a conformal map to the
``$z$-plane''. This is not the radial quantization
of~\cite{Jackiw}).
 
As far as we know, Mehta was the first to attempt to
write down a continuous one-parameter family of actions
$\calL_\theta$ such that $\calL_{\theta=0}$ is the Minkowski action
and $\calL_{\theta=\pi/2}$ is Schwinger's Euclidean action 
(although he seems 
to be unaware of Schwinger's work).
In his approach, the dependence of the Dirac matrices on the
parameter $\theta$ is not a unitary rotation, indeed it is even 
discontinuous at the midpoint $\theta=\pi/4$. An ``interpolating metric''
$(g^\theta)_{\mu\nu}$ is also introduced by hand 
such that the Euclidean action is hermitean, $SO(4)$ invariant and without
fermion doubling. 

For an application of Euclidean field theory to stochastic
processes see~\cite{Kupsch}, for a superfield formulation of Euclidean
supersymmetry see~\cite{Lukierski} and for an approach that uses zero
modes in Euclidean space to compute anomalies, see~\cite{Banerjee}. 
We shall later comment on the relation
between some of the above approaches, but first we shall introduce ours.

\section{A new Wick rotation for Dirac spinors.}

\label{sec1}

In our construction we transform only  the basic fields. Hence no extraneous
metrics $g_{\m\n}^\th$ are introduced.
The transformation induced on the Dirac matrices is unitary.
The basic observation from which we start is that for a complex vector
field $W_\mu$ one continues the combination $W_\mu^* W^\mu=
-W_0^* W_0+W_i^* W_i$ (which appears in the action)
from Minkowski to Euclidean space by giving \underline{both}
$W_\mu$ and $W_\mu^*$ the \underline{same} Wick rotation phase $e^{i\theta}$.
If $t\rightarrow e^{-i\theta}\tau$ with $\tau$ the Euclidean time coordinate 
at $\theta=\pi/2$,
\bea
W_0(t,\vec{x})&\rightarrow& 
e^{i\theta}W_4^\theta(\tau,\vec{x}),\label{four}  \\ 
W_0^*(t,\vec{x})&\rightarrow& e^{i\theta}{W_4^{\theta}}^*(\tau,\vec{x}),
\label{W}
\eea
At $\theta=\pi/2$, one then obtains ${W^\E_4}^* W^\E_4+{W^\E_i}^* W^\E_i$
where we call $W_4^{\theta=\pi/2}\equiv W^\E_4$ the fourth component of the
Euclidean field\footnote{A formal way to achieve this is to introduce 
``hyperbolic complex'' numbers of the form $a+{\bf e}b$ where $a$ and
$b$ are real and ${\bf e}$ satisfies ${\bf e}^*=-{\bf e}$, but
${\bf e}^2=+1$~\cite{Gibbons}. If one then replaces the transformation law
$W_0\rightarrow e^{i\theta}W_4(\theta)$ by $W_0\rightarrow e^{i{\bf e}\theta}
W_4(\theta)=(\cos \theta+i{\bf e}\sin \theta)W_4(\theta)$,
one finds $W_0^*W_0\rightarrow e^{2i{\bf e}\theta}
W_4^*(\theta)W_4(\theta)$ so that
at $\theta=\pi/2$ one obtains $-W_4^*W_4$.}.
We view this transformation
for vector fields
as a matrix which is diagonal with entries $(e^{i\theta},1,1,1)$. This
suggests to define a Wick rotation for a general field
as in the theory of induced 
representations: by transforming its coordinates (the orbital part) and
its indices (the spin part). It is just an accident that for vector fields
the spin matrix is diagonal, but in general the spin matrix may be more
complicated. 
   
These observations suggest that one  should consider
the following Wick rotation for a Dirac spinor
\bea
\psi(t,\vec{x})&\rightarrow&S(\theta)\psi_\theta(\tau,\vec{x})\\
\psidag(t,\vec{x})&\rightarrow&
\psidag_\theta(\tau,\vec{x})S(\theta)\label{secondeqn}
\eea
where $\psi_{\theta=\pi/2}\equiv\psi_\E$
is the Euclidean Dirac spinor.
$S(\theta)$ is a $4\times4$ matrix acting on spinor indices. 
Since  the matrix $S(\theta)$ is diagonal in the vector case,
there is at this point an ambiguity whether we should use
$S(\theta)$ or $S^\top(\theta)$ in~(\ref{secondeqn}).
At first sight one might expect that
the correct relation should involve $S^\top (\theta)$ in order
that the group property holds. However, since the group is abelian
(there is only one parameter $\theta$), also $S(\theta)$ is possible.
In fact, the correct relation involves $S(\theta)$
as in~(\ref{secondeqn}) and $S^\top (\theta)$ is inconsistent, as we shall
show.
  
In order that at $\theta=\pi/2$ the Euclidean
action contains the operator $\g_\E^\mu\d_\mu=\g^4_\E\d_4+\g_\E^k\d_k$
where\footnote{Our Euclidean signature is
$(++++)$ so Euclidean indices may be raised and lowered with 
impunity, i.e. $x^4=x_4$. We define $\g^4\equiv i\g^0$ and $\g^5
=\g^1\g^2\g^3\g^4$, similarly $\g^5_\E=\g^1_\E\g^2_\E\g^3_\E\g^4_\E$.} 
$x^4=\tau$ and $\g_\E^\mu$ satisfy the 
Clifford algebra
in~(\ref{euclid}), we define a matrix $M$ by $S(\theta)\g^4=MS\inv(\theta)$.
The matrices
\bea
\g^k(\theta)&\equiv&S\inv(\theta)\g^k S(\theta) \ = \ \g^k,\\
\g^4(\theta)&\equiv&
S\inv(\theta)\g^4 S(\theta)\ =\ \g^4\cos\theta+\g^5\sin\theta,
\eea
satisfy the Euclidean Clifford algebra. (The $i$ in 
$\partial_t=i\partial_\tau$ converts $\g^0$ into $\g^4=i\g^0$).
In order that the action at $\theta=\pi/2$ 
have an $SO(4)$ rather than $SO(3,1)$ symmetry
we require that $M$ commutes with the $SO(4)$ generators 
$[\g^\m_\E,\g^\n_\E]$.
Hence $M$ is proportional to a linear
combination of the unit matrix and $\g^5_\E$. 
It is natural to assume that the Wick rotation matrix $S(\theta)$ 
should depend only on $\g^4$ and $\g^5$ because the Wick rotation 
does not act on the space sector.
The general case $M=\a{\rm I}+\b\g^5_\E$ does not lead to an
hermitean
action and we continue with $M\sim\g^5_\E$.
The solution for the spin matrix $S(\theta)$ is given by
\be
S(\theta)=e^{\g^4\g^5\theta/2}\  ; \ \g^5=(\g^5)^\dagger\ , \ \ (\g^5)^2=1;\ 
\g^4=(\g^4)^\dagger \ , \ \ (\g^4)^2=1. 
\ee
Then 
\be
\calL_\E=-\psidag_\E(\t,\vec{x})\g^4(\g^4_\E\d_4+\g^k_\E\d_k+m)
\psi_\E(\tau,\vec{x})
\ee
where 
\bea
\g^k_\E&=&S\inv(\theta=\pi/2)\g^k S(\theta=\pi/2)\\
\g^4_\E&=&S\inv(\theta=\pi/2)\g^4 S(\theta=\pi/2).\label{5}
\eea

It is now clear that the choice $\psidag\rightarrow\psidag_\E 
S^\top$ (where $S\equiv S(\theta=\pi/2)=e^{\g^4\g^5\pi/4}$)
is not possible since defining $S^\top\g^4=(\a{\rm I}+\b\g^5_\E)S\inv$
and using $\g^5_\E=S\inv\g^5S$ leads to
$SS^\top=\a\g^4+\b\g^5\g^4$ which is not consistent as in a general
representation the right hand side is not symmetric.

Clearly, $S(\theta)$ is unitary for all real $\theta$, and 
satisfies $S(\theta)\g^4=\g^4 S\inv(\theta)$. 
Hence $\g_\E^\mu$ is hermitean. 
We note that $S(\theta)$ describes a
rotation in the ``plane'' spanned by $\g^4$ and $\g^5$. In particular,
at $\theta=\pi/2$,
\be
\g_\E^k=\g^k\ , \ \ \g_\E^4=\g^5.
\ee
Furthermore $\g_\E^5=-\g^4$. It clearly satisfies
\be
(\g^5_\E)^\dagger=\g^5_\E\ ,\ \ (\g^5_\E)^2=1\ , \ \ 
\{\g^5_\E,\g^\mu_\E\}=0
\ee
and quantities such as $(1+\g^5)/2$
appearing in an action transform into $(1+\g_\E^5)/2=(1-\g^4)/2$.  
Our Euclidean action can then finally be written as
\be
\calL_\E=\psidag_\E(\tau,\vec{x})\g^5_\E
 (\gamma^\mu_\E\partial_\mu+m)\psi_\E(\tau,\vec{x}).\label{S}
\ee
This is the hermitean action Schwinger wrote down in 1959, but for eight
component spinors. He considered real four component spinors in Minkowski
spacetime and obtained the Euclidean theory by directly rotating the
Dirac matrices. In order to still be able to impose a reality condition
he found it necessary to double the number of spinor 
components in Euclidean space.
We have given a derivation which starts from transformation rules of the 
fields, 
which is clearly more fundamental. We keep the number of components of
spinors equal to four, but since a Dirac spinor is equivalent to an eight 
component real spinor the Euclidean action we obtain is equivalent to
Schwinger's action. 
Also in Zumino's work, the $\g_\E^5$ is not explicit since he considered
massless spinors and worked with Dirac matrices $\g^\mu_{\rm Z}$
satisfying $\{\g^\mu_{\rm Z},\g^\nu_{\rm Z}\}=2\delta^{\mu\nu}$.
From our perspective $\g^\mu_{\rm Z}=i\g_\E^5\g_\E^\mu$ rather than
$\g^\mu_{\rm Z}=\g_\E^\mu$.

The procedure we have followed to obtain Euclidean spinors from their
Minkowski counterparts
boils down to the analytic continuation $t\rightarrow -i\tau$ and 
a \underline{simultaneous rotation of the spinor indices}. 
In this respect
we differ from the work of OS who continue the Minkowski theory in
time only and then construct a corresponding canonical 
Euclidean field theory. For vector fields one rotates
the indices exactly as if 
$\tau=it$ 
were part of a general coordinate transformation in the $t\t$ plane
($t^\prime=it$ yields
$W^\prime_0(t^\prime)\equiv W_4(t^\prime)=\frac{\partial t}{\partial
t^\prime}W_0(t^\prime)=-iW_0(t)$).
Requiring that the flat space vielbein field ${e_\mu}^m={\delta_\mu}^m$ 
does not
change, induces a compensating Lorentz transformation in the $4$--$5$ plane.
Our Wick rotation matrix $S(\theta)$ is just the spinor representation
of this complex finite Lorentz boost.
Further interesting questions are raised if we consider our Wick 
rotation in curved space.

\section{Dirac Spinors.}\label{ZumZumZum}

Zumino~\cite{Zumino} constructed a supersymmetric Euclidean field theory
in 1977 which follows Schwinger's approach. The aim was to incorporate 
instantons into supersymmetric theories.
The $N=2$ supersymmetric action in Euclidean space obtained by him
contains the fields $A$ (scalar), $B$ (pseudoscalar), $V_\mu$ (real vector)
and $\psi$ (Dirac spinor) just as for the Minkowski spacetime action. 
However the actions
differ in crucial parts. In Minkowski spacetime the $N=2$ supersymmetric
Yang Mills action reads
\bea{\cal L}_\M&=&-\frac{1}{4}F_{\mu\nu}^2-\frac{1}{2}(D_\mu A)^2
-\frac{1}{2}(D_\mu B)^2-\frac{1}{2}g^2(A\times B)^2 \nonumber\\&&
-\psidag\g_4\cdot\Dslash\psi-g\psidag\g^4\cdot(iA+B\gamma^5)\times\psi,
\label{buddy}
\eea
where $D_\mu A=\d _\mu A+gV_\mu \times A$, idem $D_\mu B$ and $D_\mu\psi$, 
and $F_{\mu\nu}^2=-2F_{0i}^2+F_{ij}^2$.
The Minkowski $N=2$ supersymmetry transformations are
\bea
\delta\psi&=&\Dslash A\e+i\gamma^5\Dslash B\e-
\frac{i}{2}\gamma^\mu\gamma^\nu F_{\mu\nu}\e-g(A\times B)\gamma^5\e
\label{paul}\\
\delta\psib&=&-\widebar{\e}\Dslash A+i\widebar{\e}\gamma^5\Dslash B-
\frac{i}{2}\widebar{\e}
\gamma^\mu\gamma^\nu F_{\mu\nu}+g\widebar{\e}\g^5(A\times B)
\label{george}\\
\delta A&=&\eb\psi+\psib\e\\
\delta B&=&-i(\eb\gamma^5\psi+\psib\gamma^5\e)\\
\delta V_\mu&=&-i(\eb\gamma_\mu\psi+\psib\gamma_\mu\e).
\label{mink}
\eea
where  $\psibar\equiv \psidag\g^4$.

In Euclidean space the following hermitean and
supersymmetric action was constructed\footnote{To 
match the notations in~(\ref{jogging})-(\ref{swimming}) with those 
of~\cite{Zumino}, replace 
$A_\E\rightarrow B$, $B_\E\rightarrow A$ 
(the reason for this permutation will soon become clear),
$i\g^5_\E\g^\mu_\E\rightarrow \g^\mu$, $-\g^5_\E\rightarrow\g^5$
($\g^\mu_\E\rightarrow i\g^5\g^\mu$)
and the Euclidean supersymmetry parameter
$\epsilon_\E\rightarrow-i\epsilon$.}
\bea{\cal L}_\E&=&-\frac{1}{2}{F^\E_{4i}}^{2}-\frac{1}{4}{F^\E_{ij}}^{2}
-\frac{1}{2}(D_\mu A_\E)^2
+\frac{1}{2}(D_\mu B_\E)^2+\frac{1}{2}g^2(A_\E\times B_\E)^2 \nonumber\\&&
+\psi_\E^\dagger\gamma^5_\E\cdot\Dslash^\E\psi_\E+
ig\psi^\dagger_\E\cdot(\gamma^5_\E A_\E+B_\E)\times\psi_\E,
\label{jogging}
\eea
which is invariant under the following transformation rules
\bea
\delta\psi_\E&=&\Dslash^\E A_\E\e_\E-\g^5_\E\Dslash^\E B_\E\e_\E-
\frac{i}{2}\gamma_\E^\mu\gamma_\E^\nu F_{\mu\nu}^\E\e
-ig(A_\E\times B_\E)\gamma^5_\E\e_\E
\label{neil}\\
\delta\psi_\E^\dagger&=&\e_\E^\dagger\Dslash^\E -
\e_\E^\dagger\Dslash^\E B_\E\g^5_\E-
\frac{i}{2}\e_\E^\dagger\gamma_\E^\mu\gamma_\E^\nu F_{\mu\nu}^\E
+ig\e^\dagger\gamma^5_\E(A_\E\times B_\E)
\label{tom}\\
\delta A_\E&=&-\e_\E^\dagger\g^5_\E\psi_\E-\psi_\E^\dagger\g^5_\E\e_\E\\
\delta B_\E&=&\e_\E^\dagger\psi_\E+\psi_\E^\dagger\e_\E\\
\delta V^\E_\mu&=&i(\e_\E^\dagger\gamma^5_\E\gamma_\mu^\E\psi_\E+
\psi^\dagger_\E\gamma^5_\E\gamma_\mu^\E\e_\E).
\label{swimming}
\eea
Note that the transformations $\delta A_\E$, $\delta B_\E$ and
$\delta V^\E_\mu$ are all hermitean, while $\delta \psi_\E$ and $\delta 
\psidag_\E$ are related by hermitean conjugation.

We claim that this result follows unambiguously from our Wick rotation.
To see this  we make the following transformations in the 
Minkowski action (we give only the finite transformations but one may 
reinstate the Wick rotation angle $\theta$ as outlined above to obtain
a continuous Wick rotation)
\bea
\psi\rightarrow e^{\gamma^4\gamma^5\pi/4}\psi_\E &;&
\psi^\dagger\rightarrow\psi_\E^\dagger e^{\gamma^4\gamma^5\pi/4}\ ;
\nonumber\\
V_\mu=(V_0,\vec{V})_\mu\rightarrow (iV_4^\E,\vec{V}^\E)_\mu &;&
d^4x\rightarrow-id^4x\  ;\nonumber\\
A\rightarrow A_\E &;& B\rightarrow iB_\E\ .\label{trans}
\eea
The $i$ in the rule for $V_0$ is the  $i$ from equation~(\ref{W}),
but the $i$ in the rule for the pseudoscalar $B$ needs a short explanation. 
If a pseudoscalar is represented in Minkowski spacetime in the form 
$\epsilon^{\mu\nu\rho\sigma}\d_\mu\phi_1\d_\nu\phi_2
\d_\rho\phi_3\d_\sigma\phi_4$, then under the Wick rotation 
$t\rightarrow-i\tau$, the field $B$ goes over into $iB_\E$ since only fields,
not constants such as $\epsilon^{\mu\nu\rho\sigma}$ are transformed in
our approach.

It is straightforward to check that the substitution of the 
transformations~(\ref{trans})
into the Minkowski action~(\ref{buddy}) yield the Euclidean 
action~(\ref{jogging}).
Under the same Wick rotation rules, the Minkowski supersymmetry 
transformations go over to those of the Euclidean theory provided
we at the same time also replace 
$\epsilon\rightarrow e^{\gamma^4\gamma^5\pi/4}\epsilon_\E$ and
$\epsilon^\dagger\rightarrow \epsilon^\dagger_\E e^{\gamma^4\gamma^5\pi/4}$.
This rule is clearly compatible with the Wick rotation rules of $\psi$ and 
$\psidag$ since the $N=2$ supersymmetry parameter $\epsilon$ itself is
a Dirac spinor. Further we note that our Wick rotation is consistent
with the hermiticity assignments of the continued Euclidean fields. 
This is a non-trivial statement, consider the supersymmetry 
transformation rule $\delta\psi=M\epsilon$ for some matrix $M$. 
To obtain
the rule for $\delta \psidag_\E$ one can proceed in two ways. Either one
first constructs $\delta \psi_\E$ and then takes the hermitean conjugate,
or one first takes the hermitean conjugate of $\delta \psi$ and then
continues. 
In the latter case one should use~(\ref{secondeqn}). Hence
\bea
\delta\psi=M\e\Rightarrow S\delta\psi_\E=(M)_\E S\epsilon_\E\Rightarrow
\delta\psi_\E=S\inv(M)_\E S\e_\E \nonumber\\ 
\Rightarrow\delta\psidag_\E=\epsilon_\E^\dagger S
\inv(M_E)^\dagger S,\\
\delta\psi=M\e\Rightarrow\delta \psidag=\epsilon^\dagger M^\dagger\Rightarrow
\delta\psidag_\E S=\e_\E^\dagger S(M^\dagger)_\E \nonumber\\
\Rightarrow
\delta\psidag_\E=\epsilon^\dagger_\E S(M^\dagger)_\E S\inv
\eea
(where $S\equiv e^{\gamma^4\gamma^5\pi/4}=S(\theta=\pi/2)$). 
Since the matrix $M$ may depend on fields and/or derivatives its
continuation $M_\E$ is in general not equal to $M$.
Consistency
requires
\be
S\inv(M_\E)^\dagger S=S (M^\dagger)_\E S^{-1}.\label{consistent}
\ee
One may check that the matrix $M$ of the Minkowski supersymmetry
transformation rule $\delta \psi$ in~(\ref{paul}) 
satisfies this consistency condition.

It is interesting to note that just as non-compact
$SO(3,1)$ Minkowski spacetime Lorentz transformations 
become compact $SO(4)$ rotations in Euclidean space, in contradistinction
the compact axial $U(1)$ transformations of the Minkowski theory
become non-compact scale transformations in the
Euclidean case~\cite{Zumino}\cite{Mehta}.

The requirement of hermiticity of the Euclidean action is
less justified than the property which Osterwalder and Schrader 
(and Nicolai) imposed, namely that the Greens functions in Minkowski
and Euclidean space are each other's analytic continuation.
We have shown that our approach happens to satisfy \underline{both}
requirements simultaneously for Dirac spinors. The way we have achieved
this is by not only continuing analytically the time coordinate as in OS,
but also rotating spinor indices (with the matrix $S$). 
Finally let us observe that Zumino's Euclidean pseudoscalar kinetic action
$+1/2(\d_\mu B)^2$ is not damped in the Euclidean path integral
(recall that $\exp{i\int d^4x \calL_\M}\rightarrow\exp{\int d^4x \calL_\E}$).
In this light one may prefer to trade the requirement of hermiticity
(of the action and supersymmetry transformations) for damping by 
introducing a field $B_\E^\prime=iB_\E$.

\section{Majorana spinors.}\label{Maj1}

Although we now have obtained a satisfactory treatment for Dirac spinors,
the fundamental supersymmetry models in Minkowski spacetime
are $N=1$ supersymmetric with Majorana spinors. The question now arises
what our approach yields in this case.

It is well known that the existence of
Majorana spinors depends on
the dimension and metric signature of spacetime and on whether one considers
the massive or massless case~\cite{PvN}. 
Already in 1959  Schwinger~\cite{Schwinger}
observed that no massive Majorana spinors can exist in 
four dimensional Euclidean space if one imposes a reality
condition on the spinors
themselves and hermiticity of the action.
Within a path integral approach, Nicolai~\cite{Nicolai}
obtained an action depending only on a single spinor $\psi$ (and not
$\psidag$) without any reality condition,
\be
\calL_\E({\rm Maj})=-\frac{1}{2}\psitr_\E C (\gamma^\mu_\E\partial_\mu+
m)\psi_\E
\ee
where $C$ is the Minkowski conjugation matrix and his Dirac matrices 
$\g_\E^\mu=(\g^1,\g^2,\g^3,\g^4)$ undergo no rotation. 
In this framework, he is still able to write down a supersymmetric action
since, within an $N=1$ supersymmetric model, the reality of $\psi_\E$ is not
needed to verify invariance under supersymmetry transformations. Further,
the various Fierz identities required are unchanged since the charge
conjugation matrix $C$ is still that of the Minkowski theory.

In our approach we begin with the usual Minkowski action
\be
\calL_\M({\rm Maj})=-\frac{1}{2}\psitr C (\gamma^\mu\partial_\mu+
m)\psi. 
\ee
Dropping the reality constraint, we can now unambiguously 
perform our Wick rotation on $\psi$ 
as given above. We find
\bea
\calL_\E({\rm Maj})&=&-\frac{1}{2}\psitr_\E S^\top C S  S\inv
(\gamma^\mu\partial_\mu+
m)S\psi_\E\nonumber\\
&=& -\frac{1}{2}\psitr_\E C_E  (\gamma^\mu_\E\partial_\mu+m)\psi_\E
\label{maj}
\eea
and the matrix $C_\E=S^\top C S$ also satisfies the relations
$C_\E=-C_\E^\T$ and $C_\E\g_\E^\m C_\E\inv=-{\g_\E^\m}^\top$.
Formally this action is the same as that of 
Nicolai except that Nicolai performs no rotation
on spinor indices (and therefore his charge conjugation matrix
is the original Minkowski charge conjugation matrix). Neither action
is hermitean and in neither case are the classical fields 
$\psi$ and $\psidag$ related by a reality condition. 

Let us now turn our attention to the $N=1$ Wess Zumino model~\cite{Wess}
with Minkowski action
\bea
\calL_\M&=&-\frac{1}{2}\left[(\d_\mu A)^2+m^2A^2+(\d_\mu B)^2+m^2B^2+
\psitr C (\dslash+m)\psi\right]\nonumber\\
&& -mgA(A^2+B^2)-g\psitr C(A+i\g^5B)\psi-\frac{1}{2}g^2(A^2+B^2)^2,
\eea
invariant under the transformations
\bea
&&\delta A=\epstr C \psi \ \ ; \ \ \delta B=-i\epstr C \g^5 \psi \ ;\\
&&\delta \psi =(\dslash-m)(A-i\g^5B)\epsilon-g(A-i\g^5B)^2\epsilon.
\eea
Under our Wick rotation rules $\psi\rightarrow S\psi-\E$,
$A\rightarrow A_\E$,
$B\rightarrow B_\E$, 
$\epsilon\rightarrow S\epsilon_\E$ and $t\rightarrow-i\tau$
we obtain
\bea
\hspace{-.3cm}\calL_\E&\!=\!&-\frac{1}{2}\left[(\d_\mu^\E A_\E)^2
+m^2A_\E^2+(\d_\mu^\E B_\E)^2+m^2B_\E^2+
\psitr_\E C_\E (\dslash^\E+
m)\psi_\E\right]\nonumber\\
&& -mgA_\E(A_\E^2+B_\E^2)-g\psitr_\E
 \CE (A_\E+i\g^5_\E B_\E)\psi_\E
-\frac{1}{2}g^2(A_\E^2+B_\E^2)^2,
\label{vapid}\eea
invariant under
\bea
&&\delta A_\E=\epstr_\E \CE \psi_\E
 \ \ ; \ \ \delta B_\E=i\epstr_\E \CE \g^5_\E \psi_\E 
\ ;\label{joey}\\
&&\delta \psi_\E = (\dslash^\E-m)(A_\E-i\g^5_\E B_\E)
\epsilon_\E-g(A_\E-i\g^5_\E B_\E)^2
\epsilon_\E.\label{roo}
\eea
Let us make a few comments. 
(I) Since we no longer require reality we have not
transformed the pseudoscalar field $B$ with a factor $i$. We could,
however, have done so. 
(II) The real fields $A$ and $B$ are still real in Euclidean space
unlike $\psi$. Grassmann integration $D\psi$ sees no
difference between real and complex spinors, but for bosons, Gaussian
integration over real or complex scalars yields different results.
However, the supersymmetry
transformations $\delta A_\E$ and $\delta B_\E$  can no longer be real
in the absence of a reality condition.
In practical terms however, they still provide Ward identities for
Greens functions\cite{Nicolai}. 
(III) In verifying the invariance of the $N=1$ supersymmetric
action  we must use 
certain Fierz identities whose validity is unaffected 
by our Wick transformation since $\CE$ has the same properties
as in the Minkowski case.     

In summary, at the level of path integral quantization we obtain 
an action with the same essential features as the $N=1$
supersymmetric action proposed by Nicolai. Namely a non-hermitean 
action depending
on the spinor $\psi_\E$ subject to no reality condition. 
In our approach however, Minkowski and Euclidean Greens functions are related
not only by the continuation $t\rightarrow -i\tau$ but also by the rotation
$S(\theta)$ of spinor indices.

\section{Weyl spinors.}

In Minkowski spacetime one may always rewrite a Weyl spinor as a 
Majorana spinor. However, in the previous section we found it necessary
to drop the reality condition on Majorana spinors when continuing to
Euclidean space. The Berezin integration of the path integral
is not sensitive to this modification and in this way we were able
to reproduce continued Greens functions. 
We now consider the Wick rotation for Weyl spinors.
In Minkowski spacetime $\psi_\L=\frac{1}{2}(1+\g^5)\psi$ and $\psi^\dagger_\L
=\psi^\dagger\frac{1}{2}(1+\g^5)$ is the hermitean conjugate of $\psi_\L$.
As suggested by the analysis for Majorana spinors, we should first drop
this hermiticity requirement and then perform the Wick rotation as for
Dirac spinors. Therefore we consider spinors $\psi$ and $\chi^\dagger$, 
unrelated
by hermitean or complex conjugation conjugation but nethertheless
eigenstates of $\g^5$,
\bea
\psi_{\rm L,R}=\frac{1\pm\g^5}{2}\psi &;& 
\g^5\psi_{\rm L,R}=\pm\psi_{\rm L,R}\ ;\\
\chi^\dagger_{\rm L,R}=\chi^\dagger\frac{1\pm\g^5}{2} &;& 
\chi^\dagger_{\rm L,R}\g^5=\pm\chi^\dagger_{\rm L,R}\ ;
\eea
where $\psi$ and $\chi^\dagger$ are unconstrained Dirac spinors.
Transforming $\psi\rightarrow S\psi_\E$ and $\chi^\dagger\rightarrow
\chi^\dagger_\E S$ (as usual $S=e^{\g^4\g^5\pi/4}$) in accordance with
our Wick rotation rules, we obtain the transformation rules for Weyl spinors
\bea
\psi_{\rm L,R}&\rightarrow&S\frac{1\pm\g^5_\E}{2}\psi_\E,\label{x}\\
\chi^\dagger_{\rm L,R}&\rightarrow&
\chi^\dagger_\E\frac{1\mp\g^5_\E}{2}S,\label{y}
\eea
where we used $S\inv\frac{1\pm\g^5}{2}S=\frac{1\pm\g^5_\E}{2}$
(and similarly for $S$ and $S\inv$ interchanged).
Consider now the Minkowski action for a left-handed Weyl spinor
(in which we have already relaxed hermiticity by replacing $\psidag
\rightarrow\chi^\dagger$)
\be
\calL_\M({\rm Weyl, L})=-\chi_\L^\dagger\g^4\dslash \psi_\L=
-\chi^\dagger\g^4\dslash\frac{1+\g^5}{2}\psi.
\ee
Under Wick rotation we obtain 
\be
\calL_\E({\rm Weyl, L})=\chi^\dagger_\E
\g^5_\E\dslash^\E \frac{1+\g^5_\E}{2}\psi_\E
=-{{\chi^\E_\R}^\dagger}\dslash^\E{\psi^\E}_\L.
\ee
where in Euclidean space $\psi_{\L,\R}^\E=\frac{1\pm\g^5_\E}{2}\psi_\E$ 
and ${\chi_{\L,\R}^\E}^\dagger=\chi^\dagger_\E\frac{1\pm\g^5_\E}{2}$.
As in the the Majorana case this action is not (and cannot be)
hermitean. Further notice that the $\chi^\dagger_\L\leftrightarrow\psi_L$ 
coupling
of Minkowski space becomes $\chi^\dagger_\R\leftrightarrow\psi_L$ in 
Euclidean 
space~\cite{Mehta}.

Let us now apply this formalism to the $N=1$
super Yang Mills theory with Minkowski action
\be
\calL_\M=-\psidag_\L\g^4\cdot\Dslash\psi_\L-\frac{1}{4}F_{\mu\nu}^2,
\label{ABBA}
\ee
which enjoys the supersymmetry
\bea
\delta\psi_\L&=&\frac{1}{2}\g^\mu\g^\nu F_{\mu\nu}\e_\L,\label{soggy}\\
\delta\psidag_\L\g^4&=&
-\frac{1}{2}\e^\dagger_\L\g^4\g^\mu\g^\nu F_{\mu\nu},\\
\delta V_\mu&=&-\e_\L^\dagger\g^4\g_\mu\psi_\L+\psidag_\L\g^4\g_\mu\e_\L.
\label{wheeties}
\eea
Again, $D_\mu\psi=\d_\mu\psi+gV_\mu\times\psi$, the fields $\psi$ and 
$V_\mu$ are in the adjoint representation of some non-abelian gauge group
and the subscript $\L$ denotes projection by $\frac{1+\g^5}{2}$.
We now replace $\psidag$ in~(\ref{ABBA})
 by the independent spinor $\chi^\dagger$ 
and perform our Wick rotation $t\rightarrow-i\tau$,
$\psi\rightarrow S\psi_\E$, $\chi^\dagger\rightarrow\chi_\E^\dagger S$
and $V_\mu\rightarrow {\rm diag}{(i,1,1,1)_\mu}^\nu V^\E_\nu$. 
The result reads
\be
\calL_\E=-{\chi_\R^\E}^\dagger\cdot(\dslash^\E+g\Vslash^\E\times)\psi_\L^\E
-\frac{1}{4}{F_{\mu\nu}^\E}^2.\label{K}
\ee
Applying the transformation rules~(\ref{x}) and~(\ref{y}) to the 
supersymmetry 
transformations~(\ref{soggy})--(\ref{wheeties}) we obtain the 
Euclidean supersymmetry transformation rules under which the
Euclidean action~(\ref{K}) is invariant
\bea
\delta\psi_\L^\E&=&\frac{1}{2}\g^\mu_\E\g^\nu_\E F^\E_{\mu\nu}\e_\L^\E,\\
\delta{\chi_\R^\E}^\dagger\label{s1}
&=&-\frac{1}{2}{\e_\R^\E}^\dagger\g^\mu_\E\g^\nu_\E F^\E_{\mu\nu},\\
\delta V_\mu^\E&=&-({\e_\R^\E}^\dagger\g^\E_\mu\psi_\L^\E-
{\psi_\R^\E}^\dagger\g_\mu^\E\e_\L^\E).\label{s3}
\eea
As in the Majorana case, the action and the supersymmetry transformation
rules no longer respect hermiticity and the same comments as made in the 
previous section apply here.
Finally, notice also that by taking $\psi$ and 
$\chi^\dagger$ unrelated  we obviate the
need to check that the continuation procedure satisfies the consistency
condition~(\ref{consistent}).

\section{Conclusion.}

We have defined a continuous Wick rotation on Dirac spinors
$\psi$ and their hermitean conjugates by means of a Lorentz
transformation $S(\theta)$ in the Euclidean-Minkowski $t\t$ plane.
The resulting Euclidean Greens functions are not only obtained 
from their Minkowski counterparts by continuation in the time 
variable, but also by a rotation of spinor indices. In this respect our
work agrees with work of Schwinger, who, however, did not
implement this idea on the basic fields. For Majorana and Weyl
spinors however, one is forced to drop the reality/hermiticity 
condition if one wants to perform the Wick rotation and hence
our formalism, which still uses the Lorentz rotation $S$, 
now resembles the Osterwalder Schrader approach.

Our final rule for the Wick rotation on spinors is
\bea
\psi(t,\vec{x})&\rightarrow& S(\theta)\psi_\theta(\tau,\vec{x}),\\
\psidag(t,\vec{x})&\rightarrow&\psidag_\theta(\tau,\vec{x})S(\theta)
\eea
with $S(\theta)=e^{\gamma^4\gamma^5 \theta/2}$.   
Using $S(\theta)\gamma^4=\gamma^4S\inv(\theta)$, the Dirac Minkowski action
goes over into 
\be
-\psidag\gamma^4(\gamma^k\partial_k+\gamma^0\partial_0+m)\psi
\rightarrow
-\psidag_\E\gamma^4(\gamma^k_E\partial_k+\gamma^4_E\partial_4+m)\psi_E
\ee
and since we found that $\gamma^4=-\g^5_\E$, the action is clearly hermitean
and $SO(4)$ invariant. Thus the troublesome $\g^4$ is
not omitted in Euclidean space as one might expect, but rather
reinterpreted as $\g^5_\E$ in Euclidean space~\cite{Mehta}.

With the understanding of the Wick rotation provided by
our formalism one can explicitly and continuously follow how 
vertices and propagators in interacting field theories change
as one rotates from the Minkowski to the Euclidean case. This has
practical advantages, for example it explains which terms with factors of
$\g^5$ and axial vector fields in regulators for anomalies
acquire factors of $i$. Another useful application is supersymmetry;
our rules automatically generate $N=1$ and $N=2$ theories
in Euclidean space whenever they exist in Minkowski space.

Also, another obvious open question is how to generalize
our Wick rotation to odd dimensions in which there is no ``$\g^5$''.
One possibility is to double the Dirac matrices, which suggests that the
cases in which one is able to operate with undoubled spinor
degrees of freedom in Euclidean space are the exception rather than the
norm~\cite{Wally}.

In this article we have discussed only the path integral
aspects of the Wick rotation. The Osterwalder Schrader approach
is entirely based on a canonical approach, and our Wick rotation
can also be formulated in a canonical way, but this will be published
elsewhere~\cite{Wally}. 
There we also discuss the
Osterwalder Schrader reflection positivity axiom
which is the main ingredient
for establishing the existence of a positive semi-definite self-adjoint 
Hamiltonian\footnote{In more detail, this axiom requires
that there exists an anti-linear mapping $\Theta$
which transforms an arbitrary function $F$ of the fields 
at positive times into a function $\Theta F$ of fields at negative
times such that $\langle \theta F F\rangle$ is semi-positive.
If such a mapping exists one can define a Hilbert space
with positive norm and a positive transfer matrix. The choice of time
axis with respect to which time is defined is arbitrary and breaks
the four-dimensional symmetry but the results do not depend on the 
axis chosen~\cite{Montvay}.}. 

The last comment we wish to make is perhaps the most interesting but
certainly the least understood and speculative. Both in our work and also in
that of Osterwalder Schrader, one keeps noticing five-dimensional
aspects\footnote{It may be interesting also to recall the work 
of~\cite{Stelle} in which the four-dimensional Wess Zumino
model is
obtained by dimensional reduction ``by Legendre transformation'' 
from the on-shell theory five dimensions}.
For example, in the OS canonical approach, the expansion into
creation and annihilation operators has a normalisation factor
$(k_\E^2+m^2)^{-1/2}$ in Euclidean space, as opposed to
$(\vec{k}^2+m^2)^{-1/2}$ in Minkowski spacetime. The former is clearly the
usual normalization of a five-dimensional theory at $t=0$, with four space
momenta $k_\E^\mu$. Further Schwinger introduces new fields $B(x)$
for the original fields $A(x)$ which satisfy the commutation relations
$[A(x),B(y)]=\delta^4(x-y)$, which again can be viewed as an
equal time canonical commutation relation in five dimensions.
We found a Wick rotation from Minkowski to
Euclidean space which is a five-dimensional Lorentz rotation, 
but there are many more questions and ideas yet to be developed.

\subsection*{Acknowledgements.}

It is a pleasure to thank R. Schrader, P. Breitenlohner, D. Maison, 
E. Seiler and C. Preitschopf for discussions.

\end{document}